%% file: paper.tex
\documentclass[sigplan,twocolumn, nonacm]{acmart}

\setcopyright{none}
\renewcommand\footnotetextcopyrightpermission[1]{}

\usepackage[ruled,vlined]{algorithm2e}
\usepackage{tikz}
\usetikzlibrary{arrows.meta,positioning}
\usepackage{booktabs}
\usepackage{caption}
\usepackage{enumitem}
\usepackage{mdframed}
\usepackage[table,xcdraw]{xcolor}
\usepackage{subcaption}
\usepackage{listings}
\usepackage{xcolor}
\usepackage{float}
\usepackage{lstlinebgrd}
\acmDOI{}

\acmISBN{}

\acmPrice{}

\begin{document}

\title[]{SimP: Unifying Syntax- and Semantic-Guided Techniques for Efficient Program Reduction}
\subtitle{Based on the version submitted for peer review in May 2026, with minor revisions.}

\author{Ye Xiong}
\affiliation{
  \institution{Institute of Software, Chinese Academy of Sciences}
  \city{Beijing}
  \country{China}
}
\email{xiangyug@cs.washington.edu}

\author{Xiangyu Gao}
\affiliation{
  \institution{University of Washington}
  \city{Seattle}
  \state{WA}
  \country{USA}
}
\email{xiangyug@cs.washington.edu}

\author{Jocelyn Qiaochu Chen}
\affiliation{
  \institution{University of Alberta}
  \city{Edmonton}
  \state{Alberta}
  \country{Canada}
}
\email{jocelyn.chen@ualberta.ca}

\author{Mingyu Li}
\affiliation{
  \institution{Institute of Software, Chinese Academy of Sciences}
  \city{Beijing}
  \country{China}
}
\email{limingyu@ios.ac.cn}

\author{Haibo Chen}
\affiliation{
  \institution{Institute of Software, Chinese Academy of Sciences}
  \city{Beijing}
  \country{China}
}
\email{haibochen@ios.ac.cn}

\input{definitions}

\begin{abstract}

Compiler bugs are pervasive in modern compiler systems,
but the test programs that trigger them are often too large for practical debugging.
Program reduction addresses this by minimizing test program size while preserving the original bug-triggering behavior.
Existing approaches mainly rely on syntax-guided, rule-based deletion strategies that iteratively remove parts of the program in a trial-and-error manner.
While effective in reduction quality, these approaches suffer from slow reduction speed.

This paper presents \sysname{}, a program reduction framework that combines traditional reduction with LLM-based syntax- and semantic-guided reduction. 
\sysname{} leverages customized prompt design 
to guide the reduction process. 
\sysname{} synergistically combines rule-based and LLM-based reduction stages to optimize the reduction performance. 
The results show that \sysname{} improves reduction efficiency while achieving comparable reduction quality, with negligible LLM monetary cost.
\end{abstract}

\maketitle

\input{Intro}
\input{Background}

\input{Motivation}
\input{Design}
\input{Evaluation}
\input{Related}
\input{Conclusion}

\bibliographystyle{ACM-Reference-Format}
\bibliography{reference}



\end{document}

%% file: definitions.tex
\newcommand{\cut}[1]{}
\newcommand{\eg}{e.g., }
\newcommand{\ie}{i.e., }
\newcommand{\etal}{{et al.}\xspace}
\newcommand{\vs}{{vs.} }
\newcommand{\etc}{{etc.}\xspace}
\newenvironment{parafont}{\fontfamily{ptm}\selectfont}{}
\newcommand{\Para}[1]{\par\vspace{1pt}\noindent\begin{parafont}\textbf{\textit{#1}}\end{parafont}}

\newcommand{\Sec}[1]{\S\ref{sec:#1}}
\newcommand{\Fig}[1]{Fig.\ref{fig:#1}}
\newcommand{\ct}{\small \tt }
\newcommand{\nop}[1]{}

\newcommand{\xiangyu}[1]{\textcolor{blue}{[XG: #1]}}
\newcommand{\jocelyn}[1]{\textcolor{purple}{[JC: #1]}}
\newcommand{\xiongye}[1]{\textcolor{orange}{[Ye: #1]}}
\newcommand{\fixit}[1]{{\color{red}{#1}}}

\newcommand{\squishlist}{
   \begin{list}{$\bullet$}
    { \setlength{\itemsep}{0pt}      \setlength{\parsep}{3pt}
      \setlength{\topsep}{3pt}       \setlength{\partopsep}{0pt}
      \setlength{\leftmargin}{3.5mm} \setlength{\labelwidth}{1em}
      \setlength{\labelsep}{0.5em} } }

\newcommand{\squishend}{
    \end{list}  }
\newcommand{\subsubsubsection}[1]{\paragraph{#1}}
\newcommand{\sysname}{SimP}

%% file: Intro.tex
\section{Introduction}
\label{sec:intro}

Bugs are pervasive and detrimental in production compilers.
A single compiler bug can cause the generated executable to exhibit behavior that deviates from the semantics of the input program, potentially leading to serious system failures.
Detecting and fixing these bugs is essential to the performance and reliability of the compiler.
Despite decades of engineering effort, fixing compiler bugs remains an ongoing process: widely used compilers across domains (e.g., GCC~\cite{gcc}, TVM~\cite{tvm}, P4~\cite{p4}) continue to receive a steady stream of bug-fixing patches.

Existing techniques for compiler validation broadly fall into two categories: formal methods and testing-based approaches.
Formal verification and translation validation techniques~\cite{alive2} aim to prove compiler correctness, while testing-based approaches (e.g., random program generation~\cite{csmith, yarpgen} and equivalence-based testing~\cite{EMI}) are more widely adopted in practice.
The latter generate test programs and run them through the compiler: when a program triggers an incorrect compilation outcome (e.g., a crash or miscompilation), it indicates the presence of a bug.
While effective at exposing bugs, these techniques do not control the size or complexity of the programs they generate.
As a result, bug-triggering test programs are often large and complex: tools such as Csmith~\cite{csmith} and YARPGen~\cite{yarpgen} routinely produce programs with hundreds to thousands of lines of code, which are difficult for developers to analyze directly.

Despite their size, much of the code in these test programs is irrelevant to triggering the bug.
Program reduction exploits this fact: it automatically simplifies a large test program into a much smaller version that still triggers the same bug and is manageable for human developers.

\begin{figure}[t]
    \centering
    \includegraphics[width=0.9\textwidth, trim={0 500 300 0}, clip]{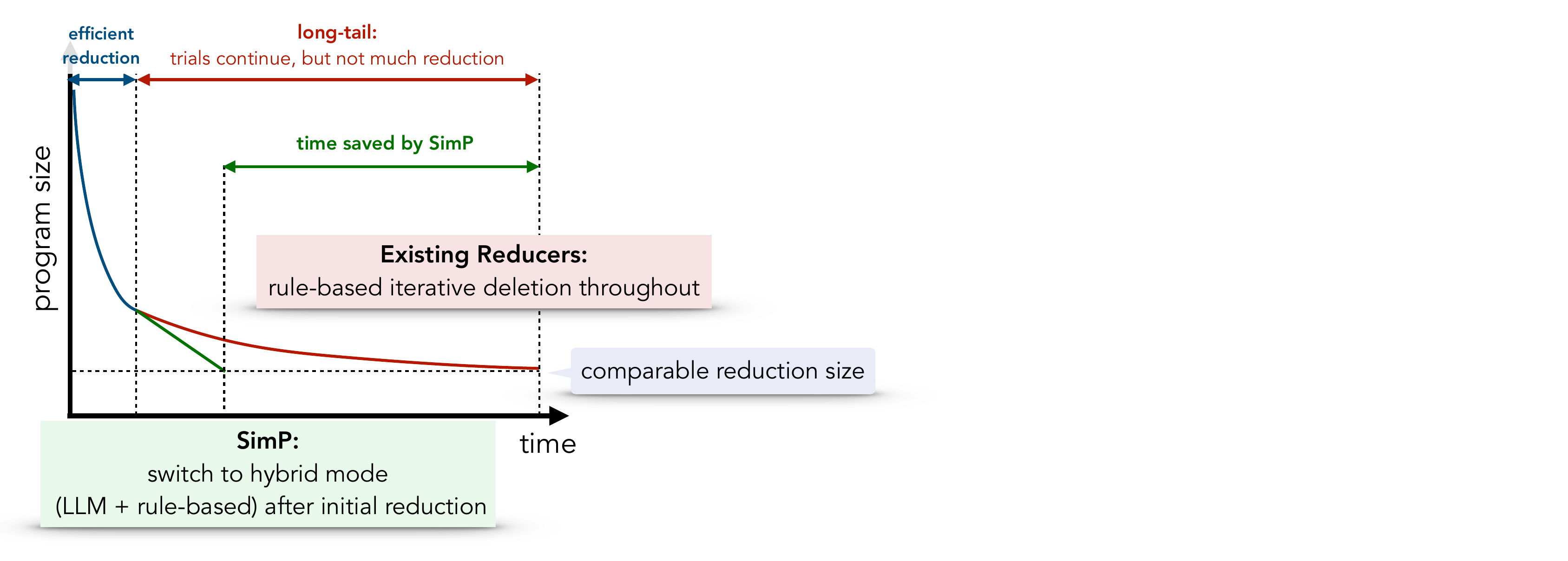}
    \vspace{-0.1in}
    \caption{State-of-the-art syntax-guided reducers can quickly remove bug-irrelevant code in the early stage, but their reduction speed gradually slows down later.
    Our approach (called \sysname{}) reuses the efficient initial reduction of existing approaches until the long-tail is detected.
    It then switches to a hybrid mode that combines LLM-guided semantic and syntactic reasoning with rule-based techniques to continue reduction.
    This avoids the long-tail behavior and improves reduction efficiency while achieving comparable reduction size.}
    \label{fig:DelvsGen}
    \vspace{-1em}
\end{figure}

State-of-the-art program reduction techniques can be broadly divided into two categories: language-generic and language-specific reducers.
Language-generic reducers (e.g., Perses~\cite{perses} and HDD~\cite{hdd}) are applicable across many programming languages, but lack language-specific techniques for deep simplification.
Language-specific reducers, such as C-Reduce~\cite{creduce} and ddSMT~\cite{ddSMT2}, leverage language-specific syntax and semantics to achieve better reduction quality, but do not generalize across languages.
Recent work such as LPR~\cite{LPR} uses large language models to enable language-aware reduction across multiple languages.

All these tools belong to the \textit{syntax-guided program reduction} category, where the reducer follows a trial-and-error process: it removes parts of the program based on its syntactic structure (e.g., abstract syntax tree), tests the validity of the reduced program (e.g., syntactic correctness and preservation of bug-triggering behavior), and then moves on to the next iteration.
However, syntax-guided reducers are slow in practice.
On publicly available benchmarks, we observe that a representative reducer, LPR, requires more than 1 hour to complete reduction in over 80\% of the cases, with over 25\% taking more than 7 hours.
Early in the reduction process, these reducers efficiently delete bug-irrelevant code from the input program.
However, as the program becomes smaller, a growing fraction of candidate transformations fail to trigger the bug or become syntactically invalid, producing a long-tail in which the reducer spends most of its time making little progress (Obs. 1 of \textsection\ref{sec:motivation}).
The root cause is that syntax-guided reducers only perform local AST edits, which cannot make the global, coordinated changes the remaining program still admits.
Spending excessive time on reduction delays the reporting of bugs to compiler developers and significantly hurts development productivity.


To remedy this issue, we propose \sysname{}, a hybrid program reduction framework that combines syntax-guided and LLM-driven reduction. The two paradigms are complementary: syntax-guided reduction handles fast bulk deletion, while LLMs can make the global semantic edits that syntax-guided reducers cannot.
As is shown in Figure~\ref{fig:DelvsGen}, \sysname{} starts with the traditional syntax-guided strategy to quickly delete bug-irrelevant code, then switches to LLM+rule-based hybrid reduction mode once the long-tail appears.

Realizing this design poses two challenges: (1) when and how to switch between syntax-guided and LLM-driven reduction, and (2) how to use LLMs reliably for reduction, both in what we ask them to do and in how we cope with the instability of their outputs.

To address the first challenge, \sysname{} uses a lightweight runtime detector that triggers the switch once two conditions hold: the program has shrunk substantially relative to the original input, and recent reduction speed has dropped below a threshold---conditions that together distinguish the long-tail from transient plateaus.
To further improve the collaboration between the two paradigms, we also remove several overly time-consuming operations from the traditional reduction workflow that contribute little to the final reduction quality.

To address the second challenge, \sysname{} customizes how it uses LLMs at each phase of reduction.
We observe that naive prompting strategies often lead LLMs toward incorrect reductions (Obs. 3 in \textsection\ref{sec:motivation}), so we design phase-specific prompts: for semantic-guided LLM reduction, prompts that guide LLMs to generate reduced programs from scratch by analyzing the root cause of the bug; for syntax-guided LLM mutation, prompts that transform the program into reduction-friendly forms that the syntax-guided reducer can shrink further.
To handle the instability of LLM outputs, \sysname{} adopts a best-of-\(N\) generation strategy: it produces multiple reduction candidates and selects the best one for the next phase.

We evaluate \sysname{} against two state-of-the-art program reducers, LPR~\cite{LPR} and Perses~\cite{perses}, on a collection of open-source and real-world compiler bug benchmarks (\textsection\ref{sec:eval}).
The results show that \sysname{} significantly improves reduction efficiency, achieving 1.75$\times$ speedup on average and up to 2.55$\times$ speedup in the best case, while maintaining reduction quality comparable to other approaches.
On average, \sysname{} saves about one hour of reduction time per benchmark.
The monetary cost of using LLMs in \sysname{} is small (less than \$0.5 per task on average), and is far outweighed by the time saved.
Our ablation study further shows that both the hybrid interaction between rule-based and LLM-based stages and the customized prompt design are essential to these improvements.

Our contributions can be summarized as follows:
\squishlist
\item \textbf{A hybrid syntax- and semantic-guided program reduction framework.}
We build \sysname{}, a hybrid program reduction framework that combines traditional rule-based, syntax-guided deletion with LLM-based syntax- and semantic-guided reduction.
\sysname{} alternates between rule-based and LLM-based stages to optimize reduction performance.
\item \textbf{Customized prompt design for program reduction.} 
We design customized prompts with human insights incorporated to guide LLMs toward preserving bug-triggering semantics while making more effective reduction decisions. 
\item \textbf{Empirical improvement in reduction efficiency.} 
Evaluation shows that on real-world compiler bug benchmarks, \sysname{} reduces reduction time while achieving reduction quality comparable to existing approaches.
\squishend

%% file: Background.tex
\section{Background}
\label{sec:background}

\subsection{Compiler bug detection}
Despite decades of engineering effort, modern compilers such as \texttt{GCC}~\cite{gcc}, \texttt{LLVM}~\cite{llvm}, \texttt{TVM}~\cite{tvm}, and domain-specific compilers (e.g., P4~\cite{p4}) continue to exhibit correctness issues, and substantial effort is devoted to identifying and fixing such bugs.
To detect compiler bugs, prior work falls into two broad categories.
Testing-based approaches, such as random program generation~\cite{csmith, yarpgen, P4Testgen} and equivalence-based testing~\cite{EMI}, generate a large number of input programs and check whether the compiled outputs satisfy expected semantic properties.
Formal verification techniques~\cite{alive2}, by contrast, prove the correctness of compiler transformations directly, without relying on concrete test inputs.
In this work, we focus on testing-based detection, which surfaces concrete failing programs that compiler developers must then diagnose.
A common class of bugs surfaced this way manifests as inconsistencies in compiler behavior: the same input program may lead to different execution results under different compilation settings.
For instance, if we use \texttt{GCC} to compile the same input program using 2 different optimization levels (e.g., \texttt{-O1} vs.\ \texttt{-O2}) but the execution results are different under these two scenarios (e.g., segmentation fault vs. normal execution), it indicates the existence of compiler bugs in at least one of the optimization modes. 

\subsection{Program reduction}
Program reduction~\cite{creduce,transformationBasedCompiler,javadeltadebug,hdd,ddSMT2,perses,CacheScheme}, is the process of simplifying an input program and preserving its property at the same time. 
Formally, given an input program $P$ and a property $\phi$ (e.g., triggering a compiler bug), 
program reduction aims to produce a reduced program $P'$ such that: (i) $P'$ preserves the property $\phi$, and (ii) $P'$ is smaller than $P$ according to a given size metric (e.g., lines of code or number of tokens).

In practice, automatically generated test programs are often large and complex, which makes them difficult to debug: a compiler developer must understand the program, pinpoint the code responsible for triggering the bug, and produce a fix, all of which become substantially harder as program size grows.
Program reduction addresses this challenge by stripping away code that is irrelevant to triggering the bug, yielding a minimal test case that is far more tractable for human analysis.



%% file: Motivation.tex
\section{Characterizing Program Reduction Approaches}
\label{sec:motivation}
\begin{figure}[t]
    \centering
    \includegraphics[width=\linewidth]{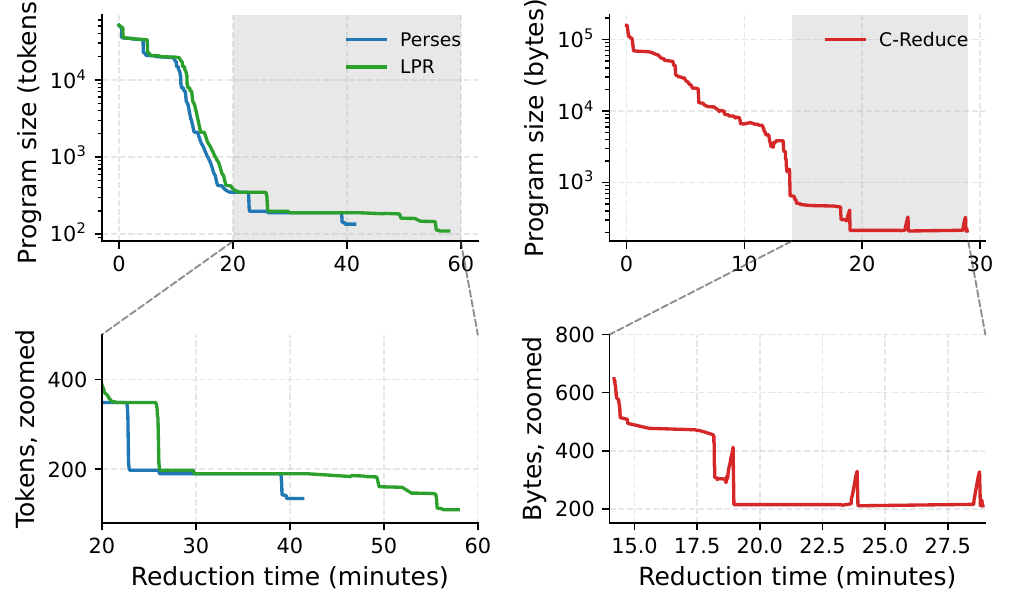}
    \vspace{-0.2in}
    \caption{Long-tail phenomenon on the gcc107176 benchmark. The top row shows the full reduction trajectory of Perses, LPR, and C-Reduce; the bottom row enlarges the shaded long-tail region of each plot.}
    \label{fig:Longtail}
    \vspace{-1em}
\end{figure}

We examine representative state-of-the-art syntax-guided reducers as well as a pure LLM-based baseline, and list three observations that expose their limitations, which motivate our design.
\Para{Observation 1: Long-tail phenomenon affects the reduction speed.}
We record the procedure of existing program reduction tools (e.g., Perses~\cite{perses}, LPR~\cite{LPR}, and C-Reduce~\cite{creduce}).
Although they can achieve high reduction speed during certain stages, they often suffer from long-tail effects throughout the reduction process.
Figure~\ref{fig:Longtail} illustrates this pattern on the gcc107176 benchmark.
In the beginning, all three reducers make rapid progress and eliminate a large number of tokens efficiently.
However, they slow down significantly in later stages, requiring substantial time to eliminate even a small number of tokens.
For instance, LPR reduces the input from roughly 51K tokens down to around 390 tokens in the first 20 minutes, but then needs another 38 minutes to shrink the remaining program from 390 to 110 tokens; the zoomed-in panels make this disparity explicit, and Perses and C-Reduce exhibit the same shape.
We observe that this phenomenon consistently appears across other benchmarks: such plateau and long-tail behaviors are common in existing syntax-guided program reducers.

\Para{Observation 2: Diminishing successful program reduction rate due to local syntactic search.}
We further look into the reasons behind this plateau and long-tail phenomenon by analyzing the intermediate output programs.
Syntax-guided reduction tools typically follow a trial-and-error approach: they reduce the current program into a smaller version, test the syntactic validity of the new program, and check whether the new program still triggers the bug or not.
If the candidate program still triggers the bug, the reducer continues reducing this new program; otherwise, it falls back to the previous version and retries other transformations.
We refer to the former as a \textit{positive program} (one that still triggers the bug) and the latter as a \textit{negative program} (one that becomes syntactically invalid or fails to reproduce the bug).

\begin{figure}[t]
    \centering
    \includegraphics[width=\linewidth]{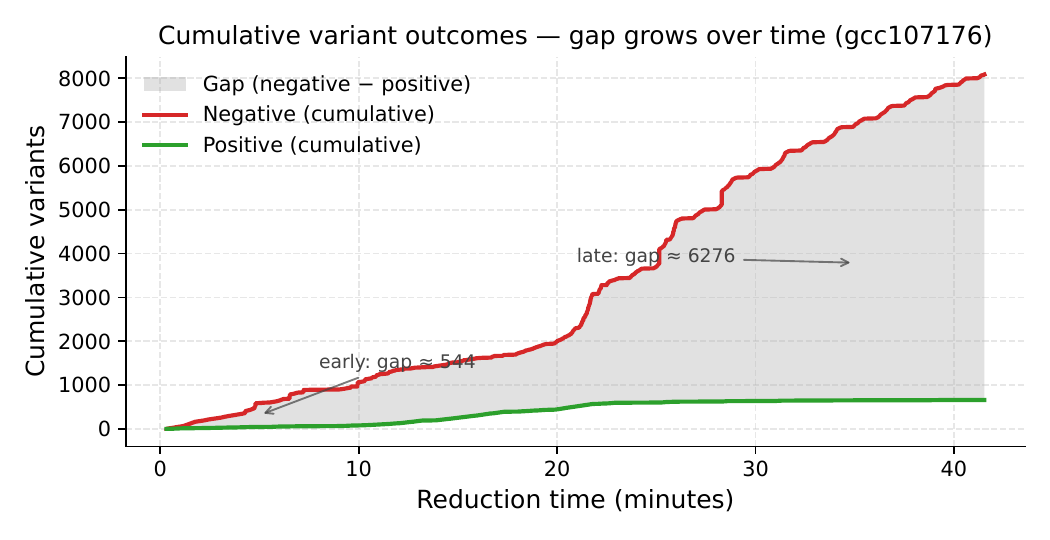}
    \vspace{-0.2in}
    \caption{Cumulative counts of negative and positive variants generated by Perses on gcc107176. The gap between the two curves grows by more than an order of magnitude between the early and late phases of reduction.}
    \label{fig:Cumulative}
\end{figure}

Figure~\ref{fig:Cumulative} shows the cumulative counts of these two types of intermediate programs in the reduction process of Perses on gcc107176.
The gap between the number of negative and positive programs is relatively small in the initial phase ($\approx$544 in the first 10 minutes), but grows significantly as the reduction process proceeds, reaching $\approx$6276 by the end.
This is because in early stages, there is ample room for reduction, so most candidate transformations succeed.
However, as the program becomes smaller, the remaining bug-triggering code fragments become increasingly difficult to simplify, and a growing fraction of candidate transformations fail.
Consequently, reducers spend more time exploring effective program variants in the later stages of reduction.

The fundamental cause is that syntax-guided reducers primarily perform local program transformations based on syntactic structures such as abstract syntax trees (ASTs).
For example, Perses categorizes AST nodes into different types (e.g., regular nodes and Kleene nodes) and iteratively removes or rewrites these nodes during reduction.
Such reductions are largely guided by local syntactic properties rather than a global understanding of the program's bug-triggering semantics.
While these local transformations can effectively eliminate large amounts of irrelevant code during the early stages of reduction, the remaining reductions in later stages often require coordinated modifications across multiple program regions (e.g., removing a function together with its callers and any types only it uses).
No single AST-local edit performs such a coordinated change, so the reducer accumulates many failed attempts before finding a sequence of moves that yields a smaller bug-preserving program -- precisely the widening gap shown in Figure~\ref{fig:Cumulative}.
Tools that build on Perses, such as LPR, inherit the same limitation.

\begin{figure}[t]
\centering

\resizebox{\columnwidth}{!}{%
\begin{tikzpicture}[
  font=\scriptsize,
  >=Stealth,
  frame/.style={draw=black!35, fill=black!2, rounded corners=2pt},
  snippet/.style={draw=black!38, rounded corners=2pt, inner sep=3pt, text width=2.65cm, minimum height=1.68cm, align=left},
  prompt/.style={draw=black!38, fill=black!5, rounded corners=2pt, inner sep=0pt, align=center},
  humanbadge/.style={draw=black!35, fill=white, circle, minimum size=0.30cm, inner sep=0pt},
  llmicon/.style={draw=black!40, fill=white, rounded corners=1pt, inner sep=1.5pt, font=\tiny\bfseries},
  callout/.style={draw=orange!65!black, fill=orange!10, rounded corners=2pt, inner xsep=4pt, inner ysep=1.5pt, align=center}
]
\node[prompt, minimum width=6.75cm, minimum height=0.36cm] (prompt) at (0,0) {};
\node[font=\scriptsize\itshape] at (0.28,0) {``Reduce the input program while preserving the bug trigger''};
\node[humanbadge] (human) at (-3.02,0) {};
\fill[black!55] (-3.02,0.055) circle (0.043);
\fill[black!55] (-3.02,-0.065) ellipse (0.090 and 0.055);

\node[frame, minimum width=7.25cm, minimum height=3.35cm, anchor=north] (loop) at (0,-0.78) {};
\node[anchor=north west, font=\scriptsize\bfseries] at (-3.45,-0.94) {Repeated hypothesis switching:};
\node[llmicon] (llm) at (3.12,-1.05) {LLM};
\foreach \x in {-0.16,-0.05,0.06,0.17} {
  \draw[black!40, line width=0.25pt] ([xshift=\x cm]llm.north) -- ++(0,0.08);
  \draw[black!40, line width=0.25pt] ([xshift=\x cm]llm.south) -- ++(0,-0.08);
}
\foreach \y in {-0.05,0.05} {
  \draw[black!40, line width=0.25pt] ([yshift=\y cm]llm.west) -- ++(-0.08,0);
  \draw[black!40, line width=0.25pt] ([yshift=\y cm]llm.east) -- ++(0.08,0);
}

\node[snippet, fill=red!7] (macro) at (-1.82,-2.3) {
  \textbf{Macro hypothesis}\\[-0.1mm]
  {\ttfamily\tiny
  \#ifdef \_\_OPTIMIZE\_\_\\
  \hspace*{0.8em}printf("3525F824");\\
  \#else\\
  \hspace*{0.8em}printf("A0552CB1");\\
  \#endif}
};

\node[snippet, fill=blue!7] (ptr) at (1.82,-2.3) {
  \textbf{Pointer hypothesis}\\[-0mm]
  {\ttfamily\tiny
  int *p = \&x, **pp = \&p;\\
  int ***ppp = \&pp;\\
  *pp = \&p3;\\
  if (p4) ****pppp = ...;}
};

\node[callout] (fail) at (0,-3.5) {bug-preserving validation fails; next trial changes track};

\draw[->] (prompt) -- (loop.north);
\draw[->, black!55] (macro.east) to[bend left=16] (ptr.west);
\draw[->, black!55] (ptr.south west) to[bend left=13] (macro.south east);
\end{tikzpicture}
}

\vspace{-2mm}
\caption{
Naive LLM-driven reduction alternates between plausible but invalid hypotheses.
The macro and pointer snippets, shown inside the switching loop, respectively fabricate optimization-level behavior and follow a pointer-nesting prototype; both fail validation because neither preserves the true bug trigger.
}

\label{fig:miscompilation_example}

\end{figure}
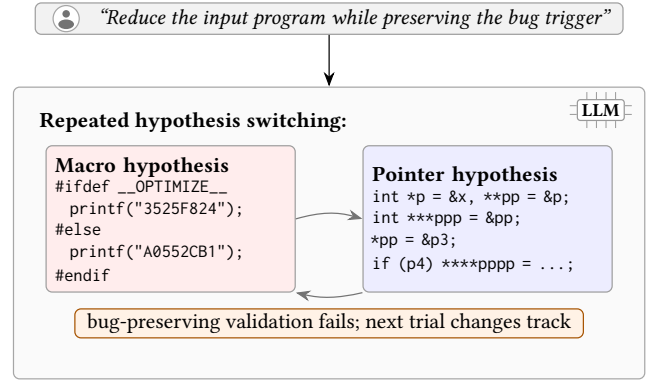

\Para{Observation 3: Pure LLM-driven program reduction is not sufficient.}
Given the wide adoption of LLMs across code-related tasks, a natural question is whether they can perform program reduction directly, bypassing the trial-and-error search used by syntax-guided tools.
We evaluate the feasibility of this approach by performing program reduction using LLMs with straightforward prompt design (e.g., please reduce the program with the bug-triggering feature preserved). 
Unfortunately, this does not achieve good reduction performance.
On the one hand, large input programs consume a substantial number of tokens, leading to high inference cost.
More seriously, the generated reduced programs are often of poor quality.

For instance, Figure~\ref{fig:miscompilation_example} shows snippets from two intermediate reduction outputs produced by an LLM for a very large Csmith-style testcase 
that exhibits different behavior under different optimization modes (\texttt{-Os} vs.\ \texttt{-O0}) in gcc-10.1.0.
The macro hypothesis shows that the LLM tries to reproduce the bug-triggering feature by using the \texttt{\_\_OPTIMIZE\_\_} macro.
However, this incorrectly attributes the root cause of the bug to the optimization flags themselves rather than to the underlying compiler miscompilation behavior.
The pointer hypothesis shows another completely different reduction trial using a hierarchy of pointers. 
Both trials fail to trigger the compiler bug. 

We check the cause of this phenomenon by analyzing LLM's reasoning. 
It shows that without guidance, LLM can, at most, identify the testcase as a miscompilation bug, but cannot provide any further diagnosis. 
The LLM reasons more like \textit{how to generate a program that behaves differently under different compilation configurations}, rather than \textit{how to generate a program that follows the trigger logic of the bug}. Hence, it either (1) observes that the program produces different outputs when compiled under different optimization levels or (2) extracts from the original testcase a specific structure that matches a compiler-bug prototype during training. 

Finally, naive prompt designs cause LLMs to frequently switch between unrelated program reduction hypotheses (e.g., pointer nesting and macro variables). 
In practice, we observe intermediate outputs similar to the macro and pointer snippets in Figure~\ref{fig:miscompilation_example} periodically appear throughout the reduction process. 
As summarized by the switching loop in Figure~\ref{fig:miscompilation_example}, the reduction goes back and forth between different reduction tracks, preventing the LLM from consistently making progress. 

\Para{Takeaway:} 
Traditional syntax-guided program reduction tools rely on iterative and heuristic-driven reduction. 
Although effective in some cases, they can suffer from a long reduction time. 
Besides, LLM-based reduction with naive prompt design also exhibits unsatisfactory performance. 
These observations motivate us to explore a hybrid program reduction approach that combines both syntax-guided and semantic-guided reduction using both traditional rule-based and emerging LLM-based techniques. 

%% file: Design.tex
\section{Our syntax- and semantic-guided reduction design}
\label{sec:design}

The design of \sysname{} is guided by a simple principle: apply different reduction strategies at different stages of reduction, so that each strategy is used where it is most effective. In the early stage, the input program is large and contains many bug-irrelevant code fragments, so \sysname{} uses a traditional syntax-guided reducer (e.g., Perses) to quickly remove large AST-level structures. Once the program enters the long-tail stage discussed in Section~\ref{sec:motivation}, local AST-level edits stall; \sysname{} then switches to two LLM-based components that complement deletion. One reasons about the program's bug-triggering semantics and synthesizes smaller bug-preserving programs from scratch. The other rewrites the program into forms that are easier for the syntax-guided reducer to shrink.

\begin{figure}[t]
    \centering
    \includegraphics[width=0.8\textwidth, trim={0 500 50 0}, clip]{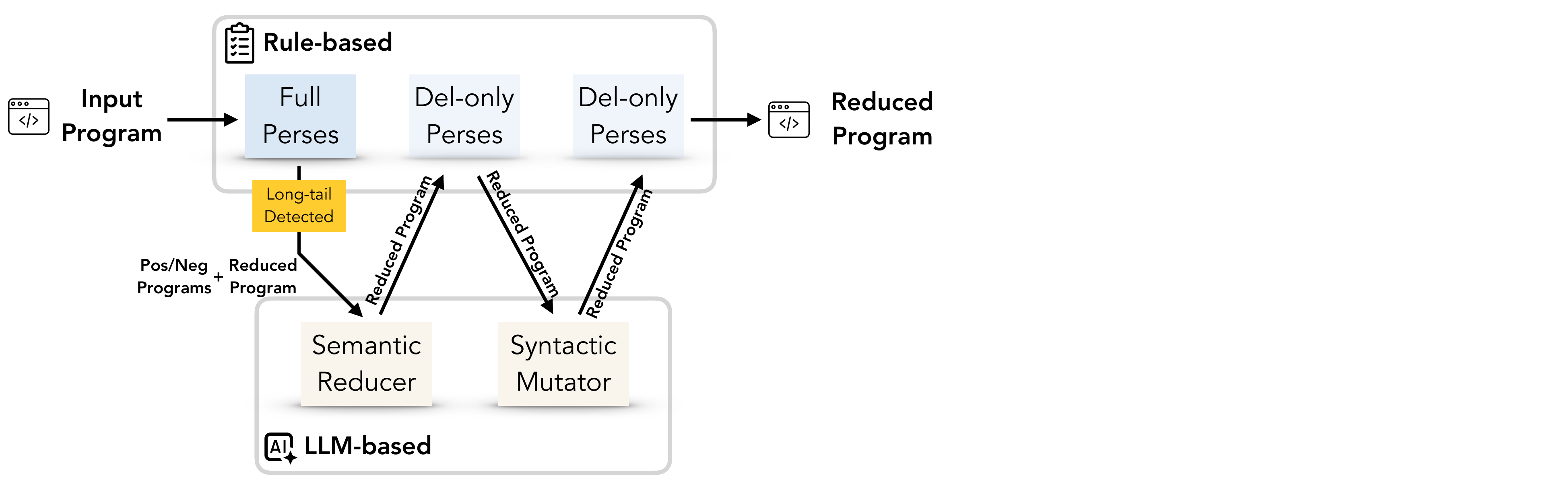}
    \vspace{-0.1in}
    \caption{
    Program reduction workflow in \sysname{}. 
    \sysname{} consists of five major stages that combine rule-based deletion and LLM-based reduction.
    The workflow starts with full Perses reduction to remove redundant bug-irrelevant code.
    Once the long-tail phenomenon is detected, \sysname{} switches to the LLM-based pipeline, which performs semantic reduction and syntactic mutation to further simplify the program.
    The generated programs are then fed back into deletion-only Perses stages for additional quick reduction. 
    }
    \label{fig:SimPWorkflow}
    \vspace{-1em}
\end{figure}

Figure~\ref{fig:SimPWorkflow} shows the resulting workflow. \sysname{} consists of three components:
(1) a \emph{syntax-guided deletion reducer} that handles early-stage reduction and includes a lightweight detector for the long-tail transition;
(2) a \emph{semantic-guided LLM reducer} that reasons about why the program triggers the bug and synthesizes smaller candidate programs that preserve the bug-triggering behavior; and
(3) a \emph{syntax-guided LLM mutator} that rewrites the program into a structurally simpler form, exposing further reduction opportunities for the deletion reducer.
The rest of this section describes each component in detail.

\subsection{Modified traditional deletion-based reduction}

Syntax-guided deletion is the cheapest reduction primitive available to \sysname{}: when the program still admits local AST edits that remove bug-irrelevant code, no other component can match its speed. \sysname{} therefore uses it in two phases of the reduction loop. The first is an initial run-to-long-tail phase that removes the bulk of bug-irrelevant code before any LLM is invoked. The second is a fast cleanup pass after each LLM stage that strips easy-to-spot fragments that LLM outputs may still contain. Both phases build on the Perses reducer~\cite{perses}, but each tailors it to its goal: the initial phase adds a runtime detector that decides when to hand off to the LLM, while the cleanup phase disables expensive rewrite operations that the LLM mutator handles more effectively.

\subsubsection{Run-to-long-tail deletion}

\sysname{} first applies Perses to perform syntax-guided reduction on the input program. As discussed earlier, syntax-guided deletion is effective during the early stage of reduction because large AST-level structures can often be removed quickly while the bug-triggering behavior is preserved. The key challenge is knowing when this strategy stops being effective: running Perses to convergence wastes time in the long-tail, but switching too early forfeits Perses's strength on bulk deletion.

To find the right switching point, \sysname{} continuously monitors the reduction progress at runtime and switches to the LLM reducer once two conditions hold: (1) the program size has been substantially reduced relative to the original input, and (2) the recent reduction speed has dropped significantly. The first condition prevents premature switching while Perses is still making bulk progress; the second catches the transition into the long-tail regime documented in \textsection\ref{sec:motivation}.

\Para{Concrete mechanism to detect long-tail.}
The main difficulty in detecting the long-tail is distinguishing it from a transient \emph{plateau}. \sysname{} addresses this with two complementary signals.
The first is an absolute token threshold $\lambda$ that guards against switching too early. While $token_t > \lambda$, where $token_t$ denotes the token count at time $t$, \sysname{} keeps Perses running regardless of its instantaneous speed: a still-large program almost certainly contains bulk irrelevant code that Perses can remove cheaply.
The second is a $score$ that captures whether reduction has structurally slowed down:
\[
\begin{aligned}
score_t &= r_t + p \cdot c_t \\
\text{where } \quad
r_t = \frac{token_t}{token_0}&,
c_t = \frac{
token_{t(1-j\%)} - token_t
}{
token_0 - token_t
}
\end{aligned}
\]
Here, $r_t$ is the remaining token ratio relative to the original program; $c_t$ is the fraction of tokens removed within the most recent elapsed-time window of length $j\% \times t$, normalized by the total reduction achieved so far; 
and $p$ is a tunable weight balancing the two terms. A small $r_t$ indicates that most removable code has already been eliminated, while a small $c_t$ indicates that recent reduction progress has slowed compared with historical progress. \sysname{} declares the long-tail phase entered (and hands off to LLM-based components) once both $token_t \le \lambda$ and $score_t \le K$ hold. 
Concrete parameters' values are mentioned in \textsection C in the appendix.









\subsubsection{Fast-and-coarse reduction}
After the LLM-based reducer (\textsection\ref{sec:llm_reducer}) and mutator (\textsection\ref{sec:llm_mutator}) return a candidate program, \sysname{} runs syntax-guided deletion again to clean up the easy-to-spot bug-irrelevant fragments that the LLM's output may still contain. This pass uses a streamlined version of Perses: we preserve only the AST-node-level deletion and simplification operations, and terminate as soon as a complete AST traversal yields no further reductions.

The original Perses additionally performs localized program rewrites to escape local minima during reduction. While effective in some cases, this exploration is expensive: it enumerates a large number of candidate variants per iteration. In \sysname{}, we disable this rewrite-based exploration and instead rely on the LLM mutator to escape local optima.

\subsection{Semantic-guided LLM reducer}
\label{sec:llm_reducer}

Once Perses enters the long-tail, the remaining edits typically require coordinated changes across multiple program regions that no single local AST edit can express (Obs. 2 in \textsection\ref{sec:motivation}). \sysname{} uses an LLM to reason about the program globally: instead of incrementally deleting fragments, the LLM identifies what makes the program trigger the bug and synthesizes a smaller program that preserves the same behavior. As is shown in Obs. 3 in \textsection\ref{sec:motivation}, however, naively prompting an LLM to do this is unreliable -- the model often switches between unrelated hypotheses about the bug's root cause. \sysname{} addresses this with three design choices: contrastive examples collected from the prior Perses run, a structured reasoning guide in the prompt, and best-of-$N$ selection across multiple sampled candidates. Algorithm~\ref{alg:semantic_reduction} summarizes the overall workflow.

\begin{algorithm}[t]
  \caption{LLM-guided semantic reduction workflow}
  \label{alg:semantic_reduction}

  \KwIn{A set of positive and negative programs $P = P^+ \cup P^-$, candidate count $N$}
  \KwOut{A reduced program $p^*$}

  $\mathcal{C} \leftarrow \emptyset$

  \ForEach{$i \in \{1, \dots, N\}$}{
      $\mathcal{C} \leftarrow \mathcal{C} \cup \{\texttt{LLMReduce}(P^+, P^-)\}$
  }

  $\mathcal{C}^{\text{valid}} \leftarrow \{p_i \in \mathcal{C} \mid \texttt{TriggersBug}(p_i)\}$

  $p^* \leftarrow \arg\min_{p_i \in \mathcal{C}^{\text{valid}}} \mathsf{sizeof}(p_i)$ // Best-of-N selection

  \Return{$p^*$}
\end{algorithm}

\subsubsection{Example collection from the prior reduction phase}
The LLM cannot identify the bug-triggering features from the program alone, since whether a fragment matters depends on the compiler and its configuration. \sysname{} therefore supplies this information indirectly through contrastive examples mined from the preceding Perses run. Each Perses iteration produces a candidate program along with its compilation result: candidates that still trigger the bug serve as \emph{positive examples}, and those that no longer do serve as \emph{negative examples}. The pairwise differences between them implicitly localize which features are responsible for triggering the bug.

To balance prompt size and information content, \sysname{} selects two positive and two negative examples generated shortly before \sysname{} switches away from Perses, when the candidates are closest in size to the program being handed off. These contrastive examples help the LLM infer which semantic features matter for this specific reduction task.

{%
\SetAlgorithmName{Procedure}{procedure}{List of Procedures}%
\begin{algorithm}[t]
  \caption{$\mathsf{LLMReduce}(P^+,P^-)$ procedure}
  \label{proc:llm_reduce_procedure}
  \DontPrintSemicolon
  \LinesNotNumbered

  \KwIn{Positive examples $P^+$, negative examples $P^-$}
  \KwOut{Reduced program $p$}

  \textbf{Step 1: Semantic Modeling}\;
  \Indp
    Analyze call chains\;
    Analyze control/data flow\;
    Infer the program's observable behavior\;
  \Indm

  \textbf{Step 2: Bug Semantic Analysis}\;
  \Indp
    Compare positive and negative examples\;
    Identify semantic differences that explain bug triggering\;
  \Indm

  \textbf{Step 3: Reduction Planning}\;
  \Indp
    Identify must-keep semantic units\;
    Remove likely-irrelevant code structures\;
    Generate reduced draft\;
  \Indm

  \textbf{Step 4: Constraint Validation}\;
  \Indp
    Preserve bug-triggering behavior\;
    Ensure compilability\;
    Avoid introducing new undefined behavior (UB)\;
  \Indm
\end{algorithm}%
}

\subsubsection{Guided semantic reasoning}
Without explicit structure, the LLM tends to jump between bug hypotheses (Observation~3) and produces unreliable reductions. \sysname{} therefore includes a structured reasoning guide in the prompt that walks the LLM through four steps before it commits to a reduced program. Procedure~\ref{proc:llm_reduce_procedure} shows the general structure; we additionally optimize the concrete prompt using DSPy~\cite{DSPy}, and the final template is available in our artifact~\cite{artifact}.

First, the LLM models the expected semantics of each positive and negative example independently. This includes identifying the call chain, summarizing the control-flow skeleton, tracking data-flow dependencies, and describing the program's observable behavior.
Second, the LLM performs bug semantic analysis by comparing positive and negative examples: it analyzes which compiler configurations expose the semantic gap, identifies code structures associated with the gap, and examines how textual differences between positive and negative examples affect program semantics.
Third, the LLM constructs a reduction plan by mapping the inferred bug semantics to concrete semantic units in the program (e.g., loop structures, call chains, aliasing patterns), and classifies each unit as \textit{must-keep}, \textit{uncertain}, or \textit{likely-irrelevant}. Based on this classification, it produces a reduced draft program as a high-confidence reproduction hypothesis. 
Take Figure~\ref{fig:LLMSemanticRed} as an example. 
Guided by its understanding of the compilation differences in the semantic level, \sysname{} identifies the \texttt{infinite loop} as the root cause of the bug.
It therefore reproduces the bug-triggering feature by preserving the \texttt{for loop} with other irrelevant components eliminated.

Finally, the LLM checks the reduced draft against a checklist of mandatory constraints set based on human experience: the draft must remain consistent with the inferred bug semantics and must be valid under the given compilation configurations. \sysname{} uses this checklist as a guardrail against obvious failure modes -- for instance, it prevents the LLM from simply replaying a previously seen positive example. 

Beyond the reduced program, \sysname{} also allows the LLM to output its reasoning trace. The trace is not consumed by downstream components, but offers useful diagnostic information for understanding why the LLM kept or removed particular program fragments.

\subsubsection{Best-of-$N$ program selection}
Even under a fixed prompt, an LLM's outputs vary across runs because of the stochastic nature of autoregressive decoding. 
\sysname{} turns this variability into an advantage by exploring multiple reduction candidates in parallel: rather than asking for a single reduced program, the prompt instructs the LLM to emit $N$ candidates (\textit{Program 1, Program 2, ..., Program N}) in a single call ($N=6$ in our implementation). 
Each candidate is then compiled and executed to check whether it still triggers the original bug; among the surviving candidates, \sysname{} selects the smallest one for the next reduction stage.

\subsection{Syntax-guided LLM mutator}
\label{sec:llm_mutator}

After a round of syntax-guided deletion, the program is often as small as local AST edits can make it but still contains structural patterns -- nested expressions, indirect calls, redundant typedefs, and the like -- that hide further reduction opportunities from the deletion-based reducer. The LLM mutator closes this gap. Unlike the semantic reducer, which synthesizes a new program from scratch, the mutator preserves the program's overall structure and applies targeted rewrites that flatten and canonicalize these patterns, transforming the program into a form that the next round of syntax-guided deletion can shrink more effectively.

\begin{algorithm}[t]
\caption{LLM-guided mutation workflow}
\label{alg:semantic_mutation}

\KwIn{Input program $p$, mutation strategies $\mathcal{M}$, regions $\mathcal{R}$}
\KwOut{A reduced program $p^*$}

$\mathcal{C} \leftarrow \emptyset$

\ForEach{$m \in \mathcal{M}$}{
    \ForEach{$r \in \mathcal{R}$}{
        // single-strategy mutation
        $\mathcal{C} \leftarrow \mathcal{C} \cup \{\texttt{LLMMutate}(p, m, r)\}$ 
    }
}
// Bug-preservation filter
$\mathcal{C}^{\text{valid}} \leftarrow \{c \in \mathcal{C} \mid \texttt{BugPreserveValidate}(c)\}$ 

// Synthesize surviving mutations
$p^* \leftarrow \texttt{LLMSynthesize}(p, \mathcal{C}^{\text{valid}})$ 

\Return{$p^*$}
\end{algorithm}

\subsubsection{Prompt design for mutation}
The prompt lists a fixed set of mutation strategies and asks the LLM to apply one strategy at a time, starting from the input program. Restricting each invocation to a single strategy keeps the resulting mutations independent of each other, which is essential for the filtering step described below: it lets \sysname{} judge which strategies are bug-preserving on this program without their effects being entangled.

Even within a single strategy, however, we observe that applying the mutation everywhere it could fire often destroys the bug-triggering feature. To mitigate this, \sysname{} restricts each mutation to a subset of the program,  increasing the probability that some regional variants can preserve the bug.

We design five categories of mutators: 

\begin{itemize}
    \item \textbf{Declaration Simplifiers.}
    These mutators simplify the declaration system by removing redundant declarations and canonicalizing aliases such as \texttt{typedef}s.

    \item \textbf{Function Inliners.}
    These mutators inline simple function calls directly into their call sites to simplify the call graph and expose more reduction opportunities.

    \item \textbf{Variable-Oriented Reducers.}
    These mutators simplify complex variable structures and reduce variable usages through techniques such as aggregate unfolding, temporary-variable inlining, parameter propagation, constant substitution, and variable merging.

    \item \textbf{Branch/Loop Reducers.}
    These mutators simplify control-flow structures by pruning unreachable branches and unrolling small fixed-trip loops, resulting in a flatter AST structure.

    \item \textbf{Expression Simplifiers.}
    These mutators simplify expressions by lifting nested expressions, reducing dereference operations, canonicalizing equivalent expressions, and simplifying arithmetic computations.
\end{itemize}

Each mutator generates three variants, yielding 15 candidate programs in total. Algorithm~\ref{alg:semantic_mutation} gives the concrete generation procedure.

Figure~\ref{fig:mutation_workflow} shows representative outputs generated by different mutators in \sysname{}.
Each mutator produces a separate candidate program that only applies the transformation strategy specified in its prompt.
For example, the \textit{Declaration Simplifier} only rewrites declaration-related constructs, while the \textit{Function Inliner} only performs local function inlining.
This design allows \sysname{} to independently explore multiple directions.

\subsubsection{Program filtering and synthesis}
\leavevmode
\Para{Bug-preservation filter.}
\sysname{} compiles and executes each of the 15 candidate programs under different compiler configurations (e.g., optimization modes) and compares their execution results. 
If the observed behaviors remain inconsistent across compiler configurations, the candidate is considered to still trigger the original compiler bug and is preserved. 
Candidates that no longer exhibit the bug-triggering behavior are filtered out. 
Only the surviving candidates are passed to the next step.

\setlength{\fboxsep}{0.7pt} 

\begin{figure}[t]
\begin{minipage}[t]{0.5\linewidth}
\begin{lstlisting}[
language=C,
basicstyle=\ttfamily\footnotesize,
escapeinside={(*@}{@*)}
]
unsigned f1(...) {...}
unsigned f2(...) {...}
unsigned f3(...) {...}
long long f4(...) {...}
... // Global Var init
void func_1(void) {
  int l0 = ...;
  char l1 = ...;
  (*@\colorbox{yellow!40}{for(g0=1; g0>=0; g0-=1);}@*)
  l0 &= f3(...);
  (*@\colorbox{yellow!40}{for (;}@*)
  (*@\colorbox{yellow!40}{  (char)f4(...)+g1;}@*)
  (*@\colorbox{yellow!40}{  g1 = (1687512307) + g1);}@*)
}
int main() { func_1(); }
\end{lstlisting}
\begin{center}
{\small (a)}
\end{center}
\end{minipage}
\hfill
\begin{minipage}[t]{0.49\linewidth}
\begin{lstlisting}[
language=C,
basicstyle=\ttfamily\footnotesize,
escapeinside={(*@}{@*)}
]
configurations:
- gcc-14.1.0 -O3 
--> (*@\colorbox{red!40}{No termination, infinite loop}@*)
- gcc-14.1.0 -O0 
--> return 0
\end{lstlisting}
\begin{center}
{\small (b)}
\end{center}
\begin{lstlisting}[
language=C,
basicstyle=\ttfamily\footnotesize,
escapeinside={(*@}{@*)}
]
... // Global Var init
long long f4(...) {...}
int main() {
  (*@\colorbox{yellow!40}{for(g0=1; g0>=0; g0-=1);}@*)
  (*@\colorbox{yellow!40}{for (;}@*)
  (*@\colorbox{yellow!40}{  (char)f4(...)+g1;}@*)
  (*@\colorbox{yellow!40}{  g1 = (1687512307) + g1);}@*)
}
\end{lstlisting}
\begin{center}
{\small (c)}
\end{center}
\end{minipage}
\caption{(a) Input bug-triggering program.
(b) Compiler behaviors: -O3 produces an infinite loop, while -O0 operates normally. LLM semantically analyzes the behavior and generates an output program (c) with only \colorbox{yellow!40}{for loops} preserved.}
\label{fig:LLMSemanticRed}
\end{figure}

\begin{figure*}[t]
\centering

\begin{minipage}[t]{0.25\textwidth}
\centering
\colorbox{green}{\small Declaration Simplifier}

\begin{lstlisting}[
language=C,
basicstyle=\ttfamily\footnotesize,
escapeinside={(*@}{@*)}
]
int printf(...);

(*@\colorbox{green!40}{long long}@*) g = 1LL;

(*@\colorbox{green!40}{long long}@*) func_3(
    (*@\colorbox{green!40}{int}@*) p14, 
    (*@\colorbox{green!40}{unsigned}@*) p15, 
    (*@\colorbox{green!40}{int}@*) p16
) {
    return p15;
}

void func_1(void) {
  for (g = 0; g < 1;)
    g=func_3(g+59, g-90, 
      g-30)+90+1;
}

int main(void) {
    func_1();
    printf("%lld\n", g);
    return 0;
}

\end{lstlisting}
\end{minipage}
\vspace{-0.2in}
\begin{minipage}[t]{0.25\textwidth}
\centering
\colorbox{green}{\small Expression Simplifier}

\begin{lstlisting}[
language=C,
basicstyle=\ttfamily\footnotesize,
escapeinside={(*@}{@*)}
]
typedef long long int64_t;
typedef int int32_t;
typedef unsigned uint32_t;
int printf(...);

int64_t g = 1LL;

int64_t func_3(
int32_t p14, 
uint32_t p15, 
int32_t p16) {
    return p15;
}

void func_1(void) {
    for(g=0; g<1;)
        g=func_3(
   g+59,g-90,g-30
)+(*@\colorbox{blue!40}{91}@*);
}

int main(void) {
    func_1();
    printf("%lld\n", g);
    return 0;
}
\end{lstlisting}
\end{minipage}
\begin{minipage}[t]{0.25\textwidth}
\centering
\colorbox{green}{\small Function Inliner}

\begin{lstlisting}[
language=C,
basicstyle=\ttfamily\footnotesize,
escapeinside={(*@}{@*)}
]
typedef long long int64_t;
typedef int int32_t;
typedef unsigned uint32_t;
int printf(...);
int64_t g = 1LL;
int64_t func_3(
    int32_t p14, 
    uint32_t p15, 
    int32_t p16
) {
    return p15;
}
int main(void) {
(*@\colorbox{yellow!40}{    for (g=0; g < 1;)}@*)
(*@\colorbox{yellow!40}{         g=func\_3(}@*)
(*@\colorbox{yellow!40}{    g+59,g-90,g-30}@*)
(*@\colorbox{yellow!40}{)+90+1;}@*)
    printf("%lld\n", g);
    return 0;
}
\end{lstlisting}
\end{minipage}
\vspace{0.1in}
\begin{minipage}[t]{0.23\textwidth}
\centering
\colorbox{red}{\small Variable-oriented Reducer}

\begin{lstlisting}[
language=C,
basicstyle=\ttfamily\footnotesize,
escapeinside={(*@}{@*)}
]
typedef long long int64_t;
typedef int int32_t;
typedef unsigned uint32_t;
int printf(...);
int64_t g = 1LL;
int64_t func_3(
    int32_t p_14, 
    int32_t p_16
) { 
    return (*@\colorbox{red!40}{g-90}@*); 
}
void func_1(void) {
  for (g = 0; g < 1;)
    g = func_3(g,g)+90+1;
}
int main() {
    func_1();
    printf("%lld\n", g);
}
\end{lstlisting}
\end{minipage}

\caption{
Intermediate mutation results for the input program in Figure~\ref{fig:input_to_merged}. 
We highlight in the code snippet that is mutated from the input program.
\colorbox{green}{Green mutated programs} preserve bug-trigger feature while \colorbox{red}{red mutated program} does not have it.}
\label{fig:mutation_workflow}

\end{figure*}

\begin{figure}[t]
\centering

\begin{minipage}[t]{0.25\textwidth}
\centering
{\small Input program}

\begin{lstlisting}[
language=C,
basicstyle=\ttfamily\footnotesize,
escapeinside={(*@}{@*)}
]
(*@\colorbox{green!40}{typedef long long int64\_t;}@*)
(*@\colorbox{green!40}{typedef int int32\_t;}@*)
(*@\colorbox{green!40}{typedef unsigned uint32\_t;}@*)
int printf(...);
int64_t g = 1LL;
int64_t func_3(
int32_t p14, 
uint32_t (*@\colorbox{red!40}{p15}@*), 
int32_t p16) {
    return (*@\colorbox{red!40}{p15}@*);
}
void (*@\colorbox{yellow!40}{func\_1}@*)(void) {
    for(g=0; g<1;)
        g=func_3(
   g+59,(*@\colorbox{red!40}{g-90}@*),g-30
)+(*@\colorbox{blue!40}{90+1}@*);
}
int main(void) {
    (*@\colorbox{yellow!40}{func\_1();}@*)
    printf("%lld\n", g);
    return 0;
}
\end{lstlisting}
\end{minipage}
\hfill
\begin{minipage}[t]{0.22\textwidth}
\centering
{\small Synthesized output program}

\begin{lstlisting}[
language=C,
basicstyle=\ttfamily\footnotesize,
escapeinside={(*@}{@*)}
]
int printf(...);

(*@\colorbox{green!40}{long long}@*) g = 1LL;

(*@\colorbox{green!40}{long long}@*) func_3(
(*@\colorbox{green!40}{int}@*) p14, 
(*@\colorbox{green!40}{unsigned}@*) p15, 
int p16) {
    return p15;
}

int main(void) {
(*@\colorbox{yellow!40}{    for (g=0; g < 1;)}@*)
(*@\colorbox{yellow!40}{         g=func\_3(}@*)
(*@\colorbox{yellow!40}{    g+59,g-90,g-30}@*)
)+(*@\colorbox{blue!40}{91}@*));
    printf("%lld\n", g);
    return 0;
}
\end{lstlisting}
\end{minipage}
\vspace{-0.1in}
\caption{Synthesized output program maintains mutations on input programs that preserve bug-triggering features. Each color represents one mutation direction in Figure~\ref{fig:mutation_workflow}. Since \colorbox{red!40}{variable-oriented reducer} fails to trigger the compiler bug, all other mutations except this one are reflected in the final synthesized output.}
\label{fig:input_to_merged}

\end{figure}

\Para{Mutation synthesis with the LLM.}
Each surviving candidate carries the effect of exactly one mutation strategy applied to one region of the program, so any single candidate captures only a partial improvement. \sysname{} therefore invokes the LLM once more to synthesize a final program that combines the useful mutations across all surviving candidates from a global view. The synthesized program is then fed into the next syntax-guided deletion stage.

Figure~\ref{fig:mutation_workflow} shows four representative mutation-specific candidates. The candidate produced by the \textit{Variable-oriented Reducer} is filtered out because it no longer preserves the original bug-triggering behavior. 
All other mutated programs can still trigger the compiler bug. 
LLM is finally invoked to synthesize these surviving transformations.  
As shown in Figure~\ref{fig:input_to_merged}, the final synthesized program preserves only the bug-preserving transformations from these candidates. 

%% file: Evaluation.tex
\section{Evaluation}
\label{sec:eval}
We evaluate \sysname{} by answering the following two questions:

\begin{enumerate}[label=Q\arabic*)]
    \item \textbf{Reduction performance:} Does \sysname{} reduce programs faster than state-of-the-art approaches without sacrificing the size of the reduced program? (\textsection\ref{sec:effectiveness})
    \item \textbf{Component contributions:} Which components of \sysname{} are responsible for these gains? (\textsection\ref{sec:ablation})
\end{enumerate}

\subsection{Experiment Setup}
\Para{Baselines.}
We compare \sysname{} against Perses~\cite{perses} and LPR~\cite{LPR}, two state-of-the-art program reducers.
Perses iteratively removes syntactic structures from the input program using grammar-aware transformations, and is widely used as a representative deletion-based reducer.
LPR builds on top of Perses and incorporates an additional LLM-based refinement pass to further improve reduction quality.
Both Perses and LPR belong to the syntax-guided program reduction category, since their reduction process is driven by iterative syntactic modifications of the program.

\Para{Benchmarks.}
We use 23 open-source compiler benchmarks from prior work~\cite{CReal, LegoFuzz}, all of which trigger \emph{miscompilation} bugs in the compiler.
Compiler bugs broadly divide into two categories: miscompilation bugs, where the compiler successfully produces an executable but the resulting program exhibits incorrect semantics, and internal compiler errors (ICEs), where the compiler itself crashes during compilation (e.g., a segmentation fault).
Prior studies have shown that ICE-triggering programs can be reduced relatively quickly~\cite{creduce}, and we confirm in our own runs that existing reducers complete on these inputs within minutes.
Miscompilation-triggering benchmarks are substantially more challenging to reduce, so we focus our evaluation on this category to stress-test \sysname{} on the harder scenario.

\Para{Metrics.}
We report three metrics: end-to-end reduction time, the size of the final reduced program, and, for the LLM-based tools, the monetary cost of API calls.
We measure program size in tokens rather than lines of code to remain consistent with the metric used by Perses~\cite{perses} and LPR~\cite{LPR}.

\Para{Environment setup.}
We run our experiments on a machine running Ubuntu 24.04 with a 120-core, 240-hyperthread Intel Xeon Platinum 8581C and 487\,GB of RAM. 
We think that any comparable Linux x86\_64 machine should reproduce the results.
For all tools that invoke LLMs, we use GPT-5.1, which we found to provide a good balance between reduction quality and API cost.

\begin{figure}[t]
    \centering
        \includegraphics[width=0.85\linewidth]{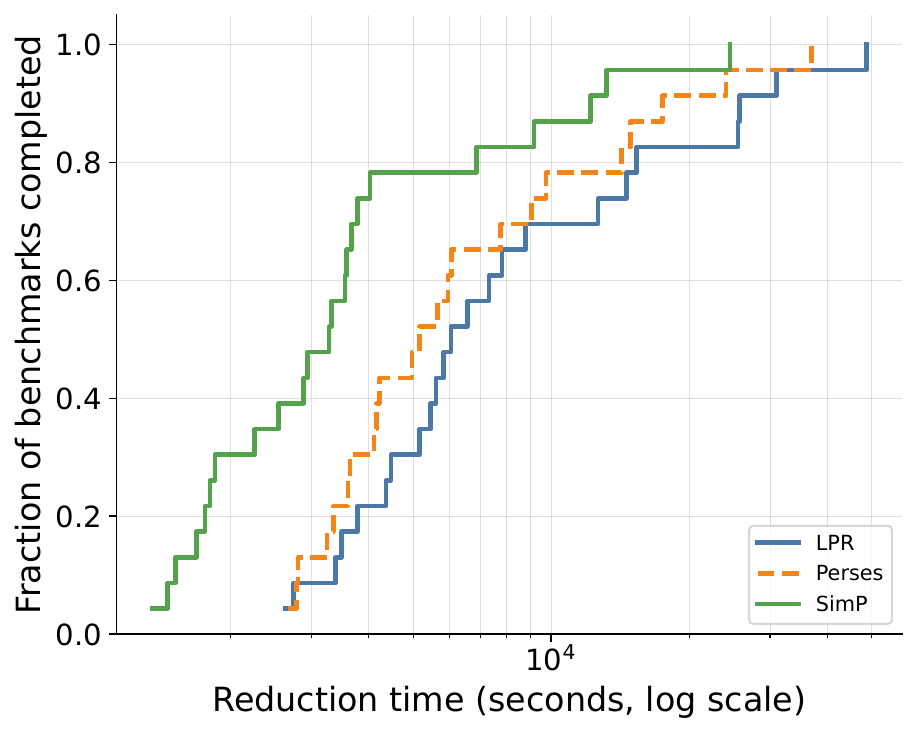}
        \vspace{-0.1in}
        \caption{
        Cumulative distribution of end-to-end reduction time across the 23 benchmarks. \sysname{}'s curve lies strictly to the left of Perses and LPR, corresponding to a 1.75$\times$ geometric-mean speedup over the better of the two.}
        \label{fig:eval_time_cdf}
\end{figure}
\begin{figure}[t]
        \includegraphics[width=0.85\linewidth]{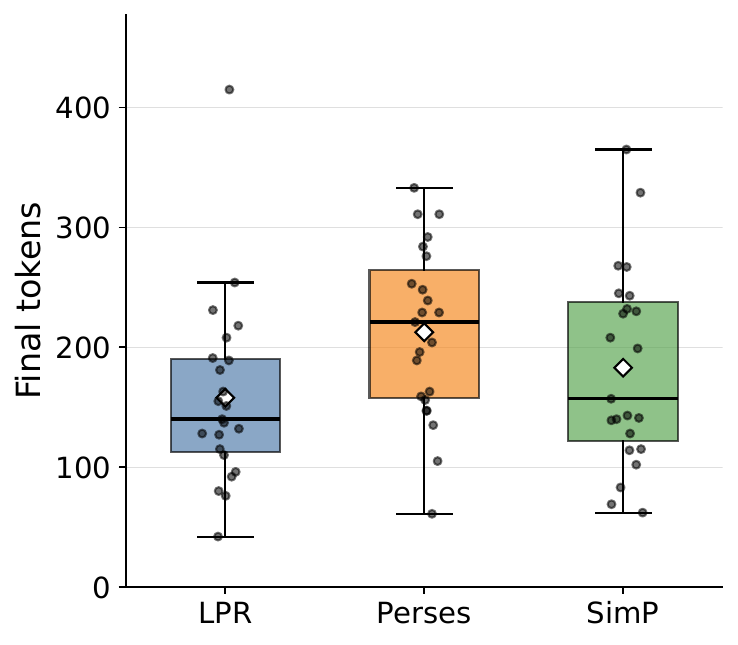}
    \vspace{-0.1in}
    \caption{
    Distribution of final reduced-program size (in tokens) across the 23 benchmarks for Perses, LPR, and \sysname{}. The three reducers produce outputs of comparable size, with no single reducer dominating across all benchmarks.}
     \label{fig:eval_final_tokens_box}
\end{figure}



\setlength{\fboxsep}{0.7pt}
\begin{figure*}[t]
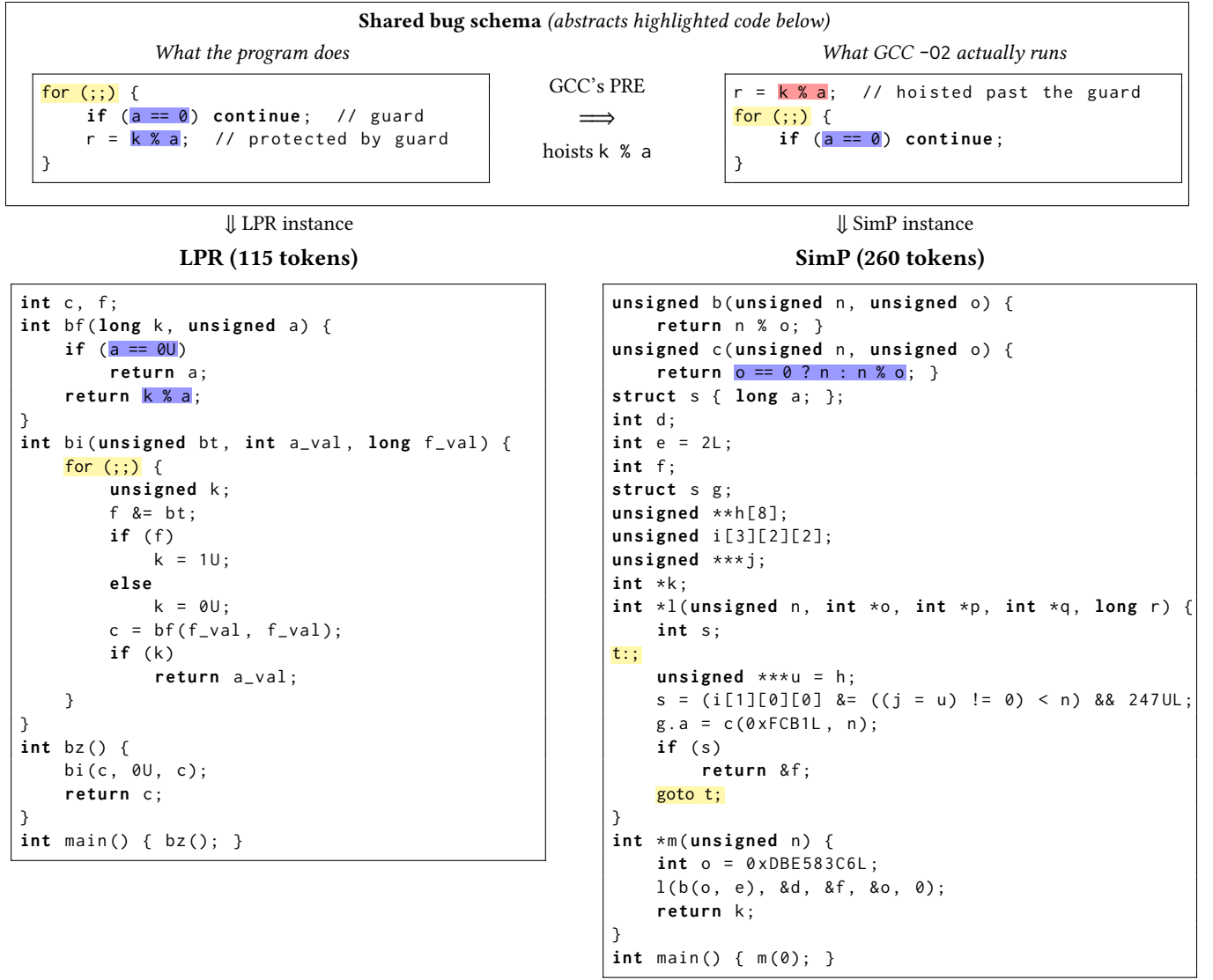

\centering

\begin{mdframed}[linewidth=0.4pt,innertopmargin=4pt,innerbottommargin=6pt,%
                 innerleftmargin=6pt,innerrightmargin=6pt]
\centering
{\small\textbf{Shared bug schema} \textit{(abstracts highlighted code below)}}
\vspace{1.5mm}

\begin{minipage}[c]{0.40\linewidth}
\centering
{\small\itshape What the program does}\\[1pt]
\begin{lstlisting}[language=C,basicstyle=\ttfamily\footnotesize,
xleftmargin=3mm,frame=single,escapeinside={(*@}{@*)}]
(*@\colorbox{yellow!40}{for (;;)}@*) {
    if ((*@\colorbox{blue!40}{a == 0}@*)) continue;  // guard
    r = (*@\colorbox{blue!40}{k \% a}@*);  // protected by guard
}
\end{lstlisting}
\end{minipage}
\hfill
\begin{minipage}[c]{0.16\linewidth}
\centering
{\small GCC's PRE}\\[2pt]
{\large$\Longrightarrow$}\\[2pt]
{\small hoists \texttt{k \% a}}
\end{minipage}
\hfill
\begin{minipage}[c]{0.40\linewidth}
\centering
{\small\itshape What GCC \texttt{-O2} actually runs}\\[1pt]
\begin{lstlisting}[language=C,basicstyle=\ttfamily\footnotesize,
xleftmargin=3mm,frame=single,escapeinside={(*@}{@*)}]
r = (*@\colorbox{red!40}{k \% a}@*);  // hoisted past the guard
(*@\colorbox{yellow!40}{for (;;)}@*) {
    if ((*@\colorbox{blue!40}{a == 0}@*)) continue;
}
\end{lstlisting}
\end{minipage}
\end{mdframed}

\vspace{1mm}
\begin{minipage}[t]{0.48\linewidth}
\centering
{\small$\Downarrow$ LPR instance}
\end{minipage}
\hfill
\begin{minipage}[t]{0.48\linewidth}
\centering
{\small$\Downarrow$ \sysname{} instance}
\end{minipage}

\vspace{1mm}

\begin{minipage}[t]{0.45\linewidth}
\centering
\textbf{LPR (115 tokens)}
\vspace{1mm}

\begin{lstlisting}[
language=C,
basicstyle=\ttfamily\footnotesize,
xleftmargin=2mm,
frame=single,
escapeinside={(*@}{@*)}
]
int c, f;
int bf(long k, unsigned a) {
    if ((*@\colorbox{blue!40}{a == 0U}@*))
        return a;
    return (*@\colorbox{blue!40}{k \% a}@*);
}
int bi(unsigned bt, int a_val, long f_val) {
    (*@\colorbox{yellow!40}{for (;;)}@*) {
        unsigned k;
        f &= bt;
        if (f)
            k = 1U;
        else
            k = 0U;
        c = bf(f_val, f_val);
        if (k)
            return a_val;
    }
}
int bz() {
    bi(c, 0U, c);
    return c;
}
int main() { bz(); }
\end{lstlisting}
\end{minipage}
\hfill
\begin{minipage}[t]{0.5\linewidth}
\centering
\textbf{\sysname{} (260 tokens)}
\vspace{1mm}

\begin{lstlisting}[
language=C,
basicstyle=\ttfamily\footnotesize,
xleftmargin=2mm,
frame=single,
escapeinside={(*@}{@*)}
]
unsigned b(unsigned n, unsigned o) {
    return n % o; }
unsigned c(unsigned n, unsigned o) {
    return (*@\colorbox{blue!40}{o == 0 ? n : n \% o}@*); }
struct s { long a; };
int d;
int e = 2L;
int f;
struct s g;
unsigned **h[8];
unsigned i[3][2][2];
unsigned ***j;
int *k;
int *l(unsigned n, int *o, int *p, int *q, long r) {
    int s;
(*@\colorbox{yellow!40}{t:;}@*)
    unsigned ***u = h;
    s = (i[1][0][0] &= ((j = u) != 0) < n) && 247UL;
    g.a = c(0xFCB1L, n);
    if (s)
        return &f;
    (*@\colorbox{yellow!40}{goto t;}@*)
}
int *m(unsigned n) {
    int o = 0xDBE583C6L;
    l(b(o, e), &d, &f, &o, 0);
    return k;
}
int main() { m(0); }
\end{lstlisting}
\end{minipage}

\vspace{-1mm}

\caption{
Qualitative comparison of reduced programs for \texttt{gcc116906}, the benchmark where \sysname{}'s output is largest relative to LPR (260 vs.\ 115 tokens).
\textbf{Top:} a compact schema abstracted from the highlighted fragments below.
\textbf{Bottom:} the concrete LPR and \sysname{} outputs that instantiate this schema; LPR builds it with plain functions and a \texttt{for} loop, while \sysname{} uses pointers, arrays, and \texttt{goto}, which accounts for its larger size.
Both outputs remain short and of similar interpretation difficulty for a human developer.
}
\label{fig:qualitative_compare}
\end{figure*}


\subsection{Reduction Performance}
\label{sec:effectiveness}
We compare \sysname{} against Perses and LPR along three axes: reduction \emph{speed} (\textsection\ref{sec:speed}), final \emph{reduction size} (\textsection\ref{sec:size}), and \emph{monetary cost} for the LLM-based tools (\textsection\ref{sec:cost}).
The first addresses the limitations raised in \textsection\ref{sec:motivation} (long reduction time); the second checks that the gains do not come at the expense of reduction quality; the third quantifies the additional cost \sysname{} incurs from invoking LLMs.
Detailed per-benchmark results are reported in Table 1 in the appendix.

\subsubsection{Reduction speed}
\label{sec:speed}
We report the end-to-end time each tool takes to complete reduction.
On average, \sysname{} finishes in 4{,}920 seconds per benchmark, compared with approximately 8{,}500 seconds for Perses and 11{,}200 seconds for LPR.
LPR is slower than Perses on 18 of 23 benchmarks because it performs an additional LLM-based refinement pass on top of Perses' output (consistent with prior work~\cite{LPR}); on the remaining five, LPR is faster.
To make the comparison conservative, we report \sysname{}'s speedup against the per-benchmark \emph{better} of Perses and LPR rather than against either tool individually.

As shown by the CDF in Figure~\ref{fig:eval_time_cdf}, \sysname{} consistently outperforms both baselines across all benchmarks.
Concretely, \sysname{} achieves a 1.75$\times$ geometric-mean speedup over the better of Perses and LPR, with a maximum speedup of 2.55$\times$.
On several benchmarks, this turns reduction time from hours into minutes.

\begin{figure}[t]
    \centering
    \includegraphics[width=0.9\linewidth]{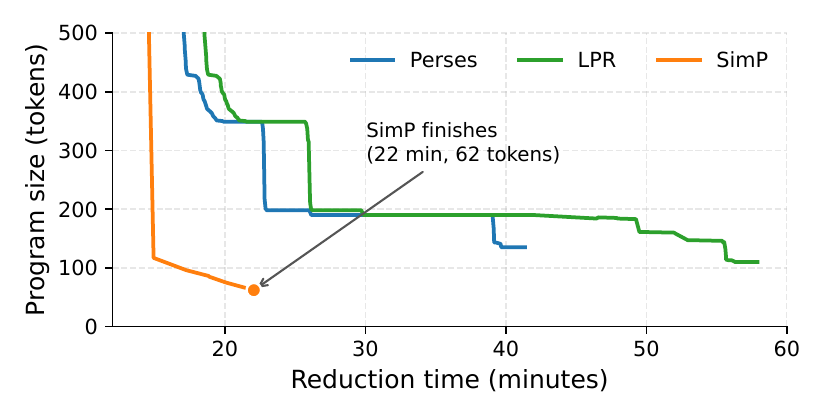}
    \vspace{-0.2in}
    \caption{
    Reduction trajectory on gcc107176, zoomed into the long-tail region of Figure~\ref{fig:Longtail}. \sysname{} converges to 62 tokens in 22 minutes; Perses and LPR are still in their plateau phase at that point and finish at 135 and 110 tokens after 41 and 58 minutes, respectively.}
    \label{fig:simp_longtail}
\end{figure}

The speedup comes from collapsing the long-tail phase that dominates the runtime of the state-of-the-art syntax-guided reducers (Obs.~1 of \textsection\ref{sec:motivation}).
Figure~\ref{fig:simp_longtail} illustrates this on gcc107176: Perses and LPR enter a plateau around 200--350 tokens and spend the next 20--40 minutes shaving off a small number of tokens at a time, whereas \sysname{}'s LLM-driven passes drop directly through this region in about two minutes and converge to a smaller final program (62 tokens) within 22 minutes.
By the time Perses and LPR finish (at 41 and 58 minutes respectively), \sysname{} has already produced a strictly smaller reduced program.
\vspace{0.05in}
\begin{mdframed}[linewidth=0.2pt, innerleftmargin=6pt, innerrightmargin=6pt]
\textbf{Summary.} \sysname{} reduces programs faster than both baselines on every benchmark, with a 1.75$\times$ geometric-mean speedup over the better of Perses and LPR. The maximum speedup is 2.55$\times$, dropping from $>$7 hours to $\approx$100 minutes.
\end{mdframed}


\subsubsection{Reduction size}
\label{sec:size}
We measure final reduction quality by the number of tokens in the output program.
Figure~\ref{fig:eval_final_tokens_box} shows the distribution of token counts across all reducers.
All three tools produce substantial reductions: every output is more than 99\% smaller than the original benchmark, and every reduced program is under 60 LoC, which is compact enough for a junior developer to read and understand the bug-triggering behavior directly.

No single reducer consistently produces the smallest output across all benchmarks.
LPR often achieves the smallest reduced programs because it performs an additional LLM-based refinement pass on top of Perses' output, while \sysname{} matches or beats both Perses and LPR on several benchmarks without such a refinement stage.

The biggest gap between LPR and \sysname{} over all benchmarks is on gcc116906 (Figure~\ref{fig:qualitative_compare}), where \sysname{}'s output is more than 2$\times$ larger than LPR's.
Figure~\ref{fig:qualitative_compare} explains why this larger output is still easy to inspect: the top panel abstracts the shared bug schema from the highlighted fragments in the two concrete programs below it.
The root cause of gcc116906 lies in GCC's Partial Redundancy Elimination (PRE) optimization, which incorrectly hoists may-trap expressions in the presence of control-flow edges that may never execute.
In LPR's output, function \texttt{bf()} contains a guarded modulo operation: \colorbox{blue!40}{$k$ \% $a$} should not execute when \colorbox{blue!40}{a == 0}.
The surrounding nonterminating loop creates the control-flow context that lets PRE hoist the modulo before the guard, eventually evaluating it with a zero divisor.
\sysname{} reproduces the same bug-triggering schema with a different construction, using pointers, arrays, and \texttt{goto}, which increases the token count.

Even on this benchmark, where LPR achieves its largest reduction advantage over \sysname{}, 
the size difference does not noticeably increase the difficulty of human inspection. 
Further closing such size gaps in an efficient way can be left for future work. 
\vspace{0.05in}
\begin{mdframed}[linewidth=0.2pt, innerleftmargin=6pt, innerrightmargin=6pt, nobreak=true]
\textbf{Summary.} \sysname{} achieves reduction quality comparable to existing approaches: all reducers shrink the original benchmarks to under 60 LoC, making the resulting programs easy for human developers to understand.
\end{mdframed}

\subsubsection{Monetary cost}
\label{sec:cost}
Both LPR and \sysname{} invoke LLMs during the reduction process. 
\sysname{} incurs a slightly higher monetary cost than LPR. However, for both tools, the total reduction cost per benchmark remains below \$0.5, and the per-benchmark cost difference is $\leq$\$0.3. 
Given that \sysname{} saves around 1h of reduction time per benchmark on average, this modest cost increase is a worthwhile trade-off.
\vspace{0.05in}
\begin{mdframed}[linewidth=0.2pt, innerleftmargin=6pt, innerrightmargin=6pt]
\textbf{Summary.} Running \sysname{} on a single benchmark costs less than \$0.5, which is worthwhile, especially given the saved reduction time.
\end{mdframed}

\begin{figure}[t]
    \centering
    \begin{subfigure}[b]{0.85\linewidth}
        \centering
        \includegraphics[width=\linewidth]{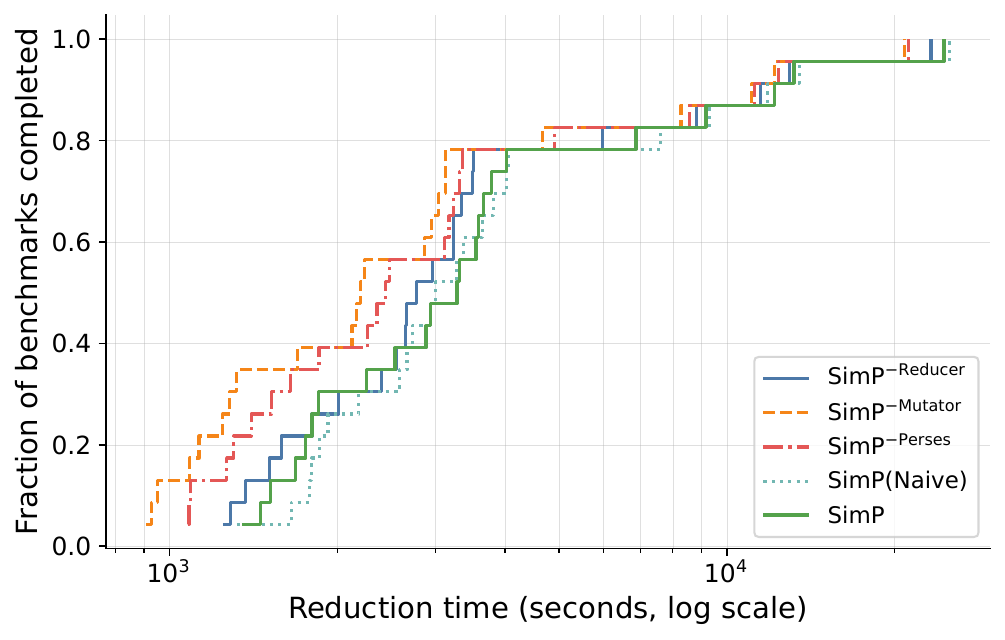}
        \vspace{-0.2in}
        \caption{Cumulative distribution of end-to-end reduction time.}
        \label{fig:ablation_reduction_time_cdf}
    \end{subfigure}


    \begin{subfigure}[b]{0.85\linewidth}
        \centering
        \includegraphics[width=\linewidth]{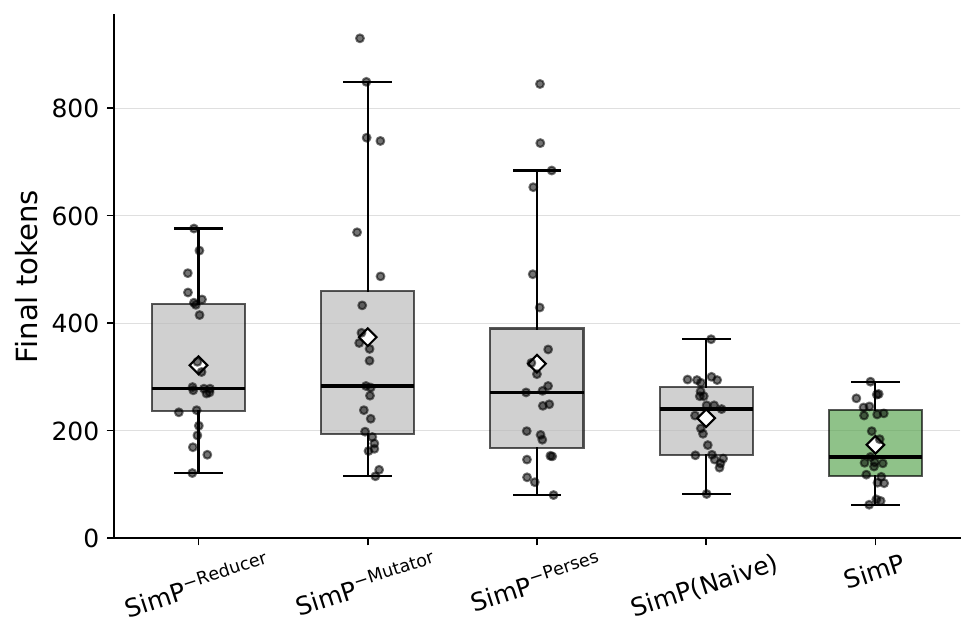}
        \caption{Distribution of final reduced-program size (in tokens).}
        \label{fig:ablation_final_tokens_boxplot}
    \end{subfigure}
    \vspace{-0.1in}
    \caption{Ablation of \sysname{}'s components across the 23 benchmarks. The syntax-guided mutator and the semantic reducer form a quality--speed trade-off, while Perses and the structured prompt design each independently improve reduction quality.}
    \label{fig:ablation}
\end{figure}

\subsection{Ablation}
\label{sec:ablation}
We isolate the contribution of each component of \sysname{} by running four ablation variants:
\squishlist
\item \textbf{Without mutator:} the syntax-guided LLM mutator disabled. 
\item \textbf{Without reducer:} the semantic-guided LLM reducer disabled. 
\item \textbf{Without Perses:} the Perses-based deletion stage disabled.
\item \textbf{Naive prompt:} the structured prompts in both LLM components are replaced by a prompt that simply asks the LLM to reduce the program while preserving the bug-triggering behavior. 
Concretely, the semantic-guided reducer is prompted only to synthesize a smaller bug-preserving program, without the four-step reasoning guide; the syntax-guided mutator is prompted only to rewrite the program into a form that is easier to reduce further, without a fixed set of mutation strategies.  
\squishend
Figures~\ref{fig:ablation_reduction_time_cdf} and~\ref{fig:ablation_final_tokens_boxplot} summarize the time and size distributions across these variants; 
per-benchmark numbers appear in Table 2 in the appendix.

The two LLM components contribute to a higher quality reduction result.
Enabling the syntax-guided mutator produces final programs 50\% smaller on average (up to 73\%), but adds about 10 minutes of reduction time per benchmark on average.
Disabling the semantic-guided reducer has the opposite effect---final programs become 47\% larger on average (up to 81\%) while reduction is only about two minutes faster---indicating that the reducer accounts for most of \sysname{}'s reduction quality at modest runtime cost.

Perses and the prompt design contribute to the reduction quality as well.
Removing Perses produces final programs 41\% larger on average (up to 68\%) while saving $\approx$10 minutes of reduction time, showing that traditional syntax-guided, AST-based reduction remains a load-bearing component of \sysname{}: the LLM-driven passes alone do not match the quality achieved when Perses handles the early-stage bulk deletion.
Replacing our structured prompts with the naive prompt produces final programs 25\% larger on average (up to 75\%) and also slows the reduction time, confirming that our prompt design improves both reduction quality and efficiency.
\begin{mdframed}[linewidth=0.2pt, innerleftmargin=6pt, innerrightmargin=6pt, nobreak=true]
\textbf{Summary.} Every component contributes measurably to \sysname{}'s overall performance. The semantic-guided LLM reducer and the syntax-guided LLM mutator form a quality-speed trade-off, while Perses and our prompt design each independently improve reduction quality.
\end{mdframed}

%% file: Related.tex
\section{Related Work}
\label{sec:related}

\Para{Compiler bug fix.}
Several works focus on detecting bugs in existing compiler implementations through static analysis, symbolic execution, differential testing, and fuzzing. 
For example, compiler fuzzers generate large numbers of random or coverage-guided test programs to trigger crashes, miscompilations, and undefined behaviors. 
\sysname{} is complementary to existing compiler bug detection and fix techniques by providing a more efficient reduction framework. 
This helps developers better understand compiler bugs and accelerate their debugging process. 



\Para{Program Reduction.}
Program reduction is important for helping compiler developers debug compiler bugs by minimizing large bug-triggering programs into compact test cases.
Existing reduction techniques mainly rely on syntax-level transformations and iterative trial-and-error validation.
These approaches often suffer from a long-tail bottleneck where further reduction becomes increasingly difficult.
\sysname{} takes a fundamentally different approach by incorporating LLM-based semantic understanding into the reduction process.
\sysname{} exhibits faster reduction speed. 

\Para{LLMs for code generation and optimization}
Large Language Models (LLMs) have been widely adopted for a variety of programming tasks, 
including code generation~\cite{codex, alphaCode, codegen}, 
program understanding~\cite{codebert, graphcodebert}, 
and optimization~\cite{llm_opt, alphaevolve}. 
Recent work has demonstrated that LLMs can effectively reason about program semantics and synthesize non-trivial programs.
These LLM models~\cite{gpt4, deepseek, gemini} are trained on large-scale code bases, learning statistical patterns and structural regularities in programs.
All these efforts belong to the domain of leveraging LLM to optimize the software engineering workflow. 
This is orthogonal to our usage of LLMs in the domain of program reduction. 

%% file: Conclusion.tex
\section{Conclusion}
\label{sec:conclusion}
Program reduction is important to help developers detect and fix hidden compiler bugs. 
Existing tools mainly rely on syntax-guided reduction approaches and suffer from long-tail reduction issues.
We build \sysname{}, a tool that consists of both syntax-guided and semantic-guided components to do program reduction. 
\sysname{} has better reduction performance than state-of-the-art approaches. 
We believe that the key insight behind \sysname{} can generalize and benefit program reduction tasks in many other domains, such as SMT solvers, database engines, and other complex software systems.